\documentclass[usenatbib]{mn2e}
\usepackage{natbib}
\usepackage{amsmath,amssymb}
\usepackage{epsfig}
\usepackage{color}

\input{macros.tex}

\title[Bringing the Galaxy's dark halo to life]
{Bringing the Galaxy's dark halo to life}

\author[Piffl, Penoyre \& Binney]{T.~Piffl$^1$\thanks{E-mail: tilmann.piffl@physics.ox.ac.uk},
Z.~Penoyre$^2$,
J.~Binney$^1$ 
\\
$^1$Rudolf Peierls Centre for Theoretical Physics, Keble Road, Oxford OX1 3NP, UK\\
$^2$Institute for Astronomy, University of Cambridge, Madingley Road, Cambridge CB3 0HA, UK\\
}

\begin{document}
\date{Draft, \today}
\pagerange{\pageref{firstpage}--\pageref{lastpage}} \pubyear{2014}
\maketitle
\label{firstpage}
\begin{abstract}
We present a new method to construct fully self-consistent equilibrium models
of multi-component disc galaxies similar to the Milky Way.  We define
distribution functions for the stellar disc and dark halo that depend on phase space
position only through action coordinates. We then use an iterative
approach to find the corresponding gravitational potential. We study the
adiabatic response of the initially spherical dark halo to the introduction
of the baryonic component and find that the halo flattens in its inner
regions with final minor-major axis ratios $q$ = 0.75 -- 0.95. The extent of
the flattening depends on the velocity structure of the halo particles with
radially biased models exhibiting a stronger response. In this latter case,
which is according to cosmological simulations the most likely one, the new density
structure resembles a ``dark disc'' superimposed on  a spherical halo. We
discuss the implications of these results for our recent estimate of the
local dark matter density.\\
The velocity distribution of the dark-matter particles near the Sun is very
non-Gaussian. All three principal velocity dispersions are boosted as the
halo contracts, and at low velocities a plateau develops in the distribution
of $v_z$. For models similar to a state-of-the-art Galaxy model we find
velocity dispersions around $180\,\mathrm{km\,s^{-1}}$ for $v_z$ and the
tangential velocity, $v_\varphi$, and 150 -- $205\,\mathrm{km\,s^{-1}}$ for
the in-plane radial velocity, $v_R$, depending on the anisotropy of the
model.
\end{abstract}

\section{Introduction} \label{sec:Intro}
It is now generally accepted that most of the rest mass in the Universe is
contained in a type of ``dark matter'' that is incapable of electromagnetic
interactions. Dark matter began to cluster gravitationally earlier than
baryonic matter because it was decoupled from the cosmic radiation bath,
which at redshifts $\gta1000$ provided abundant pressure. Hence the formation
of a galaxy starts with the accumulation of dark matter to form the dark
halo.  When and how baryons accumulated in the gravitational potential wells
of dark halos is ill understood. In the simplest picture, they simply fell in
dissipatively and formed stars near each halo's centre, but this picture is
inconsistent with the small fraction of baryons that are now in stars rather
than intergalactic gas. Several lines of evidence indicate that the release
of nuclear energy by early stars strongly attenuated the infall of baryons,
thus increasing the extent to which galaxies are dark-matter (\DM) dominated.

The least \DM\ dominated galaxies are ones with luminosities similar
to that, $L*$, of our Galaxy: dwarf galaxies are very strongly \DM\
dominated and stars play such a subsidiary role in the most massive
\DM\ halos, those of galaxy groups and clusters, that we do not even
recognise these halos as galaxies. Yet even in galaxies such as ours, it may
be argued that the baryons have played a very subordinate role.  Measurements
of the optical depth to microlensing of stars located less than $\sim10\deg$
from the Galactic centre imply that a significant majority of the mass inside
$\sim3\kpc$ is concentrated in stars \citep{BinneyEvans2001,Bissantz2004}.
But the majority of the gravitational force that holds the Sun in its orbit
comes from dark matter \citep[e.g.][hereafter P14]{Piffl2014b}, and the
dominance of dark matter increases sharply with increasing distance from the
centre. 

Our understanding of how baryons accumulated at the centres of dark halos to
form visible galaxies is limited. This is because the physics of this process is
complex and is known to involve scales $\la1\pc$ that are by far unresolved in
feasible simulations of the formation of galaxies in a cosmological context. Early
simulations produced galaxies that contained too large a fraction of the
baryons in the universe, and were much more dominated by their spheroidal
components than is our Galaxy or our near neighbour M33. In recent years it has
become clear that energy released during star formation is very efficient at
preventing the accumulation of large masses of gas and stars in the centres
of dark halos, by expelling gas that has fallen in to a halo and the products
of stellar evolution into intergalactic space, where the great majority of
the baryons must reside. The high efficiency of such ``feedback'' from star
formation is surprising and it can be reproduced in simulations only by
modifying the encoded laws in essentially arbitrary ways. 

Since we cannot trust current simulations of the behaviour of baryons during
star formation, it is worthwhile to explore simple limiting cases. One such
limiting case is that in which the baryons have accumulated in the Galaxy
gradually rather than in lumps. This scenario is favoured by four
observations. The first is that the bulge of our Galaxy bears all the
hallmarks of having formed through a thin disc of stars that formed on nearly
circular orbits in the Galactic plane developing a strong bar, which then
buckled \citep{CombesSanders1981,Raha1991,AAOmega}. The second relevant
observation is that in the last $\sim10\Gyr$ the rate of star formation in
the disc has declined by no more than a factor $\sim3$
\citep[e.g.][]{Aumer2009}. A third relevant observation is that the
star-formation rate in a disc appears to be proportional to the surface
density of cold, molecular gas \citep{Leroy2008}. A fourth observation is that
currently $\la10\%$ of the baryons in the disc are in gas.

From the indication that all but a tiny fraction of the Galaxy's stars formed
from quiescently orbiting gas in the Galactic plane, we infer that our Galaxy
has not experienced a significant merger since it started seriously forming
stars. From the sustained rate of star formation over the Galaxy's life
together with the dependence of star formation rate on cold-gas density and
the relatively small current stock of gas, we infer that the baryonic mass of
our Galaxy has accumulated over its entire life, presumably through sustained
cooling of gas from the extended corona into the disc. There is even strong
circumstantial evidence of this cooling process taking place now
\citep{Fraternali2013}.

A dark halo will respond to the quiescent accumulation of baryons in a disc
by distorting adiabatically: that is, the orbit of each \DM\ particle
will evolve in such a way that its action integrals $J_r$, $J_\phi$ and $J_z$
will be constant, with the consequence that the density of \DM\
particles in action space, $f(\vJ)$, is unaffected by the accumulation of
baryons and the resulting evolution of the galaxy's gravitational potential
$\Phi(\vx)$ \citep[e.g.][\S4.6.1]{GalacticDynamics}.  The constancy of
$f(\vJ)$ is enormously useful because the gravitational clustering of dark
matter in isolation can be simulated with confidence, so we know what dark
halos would now populate the Universe if there were no baryons. That is, from
\DM\ only simulations we can infer what $f(\vJ)$ would be in the
absence of baryons, and the adiabatic principle assures us that $f(\vJ)$ will
be unchanged by the quiescent infall of baryons in the observed quantities.
Hence the hypothesis that the baryons accumulated quiescently so the dark
matter responded adiabatically to their arrival, allows us to predict the
present structures of galaxies before we understand the complex processes
that resulted in the accumulation of baryons.

This general idea was appreciated at an early stage in the development of the
cold dark matter (CDM) cosmological paradigm \citep{Blumenthal1986}, but until
recently we have lacked the technology required to exploit it fully.
The key to its exploitation is the ability to compute the action vector
as a function $\vJ(\vx,\vv)$ of ordinary phase-space coordinates $(\vx,\vv)$,
for then it is possible to compute the \DM\ density 
\[\label{eq:rho_f}
\rho_{\rm DM}(\vx)=\int\d^3\vv\,f[\vJ(\vx,\vv)]
\]
at any location $\vx$. 

In an axisymmetric potential $\Phi$ one action,
$J_\phi=L_z$, is simply the component of angular momentum parallel to the
potential's symmetry axis, and in a spherical potential $J_z=L-|L_z|$, where
$L$ is the length of the angular-momentum vector $\vL$. \cite{Blumenthal1986}
estimated the response of a dark halo to the accumulation of baryons by
assuming that all \DM\ particles are on circular orbits, so
$J_r=0$. A more realistic estimate of halo response was obtained by
\cite{SellwoodMcGaugh2005}, who lifted the restriction $J_r=0$ to circular orbits
but retained the assumption of spherical symmetry. The latter  is awkward because
the basic picture is of baryons accumulating in a disc, which will
immediately break any pre-existing spherical symmetry of the dark halo. 

In this paper we lift the assumption of spherical symmetry, so we are able to
compute how the quiescent introduction of the disc would have deformed an
initially spherical dark halo into a more centrally concentrated, oblate
structure. We are, moreover, able to predict the full, three-dimensional
distribution of the velocities of \DM\ particles at any point in the
Galaxy, in particular at the Sun.  These distributions are triaxial and give
non-Maxwellians speed distributions.
\section{Methodology} \label{sec:Method}
Our strategy is to adopt \df s $f(\vJ)$ for the Galaxy's stellar disc and
dark halo and to solve for the gravitational potential $\Phi(\vx)$ that these
self-consistently generate in the presence of given density distributions for
the bulge and gas disc. We now describe our algorithm for the determination
of $\Phi(\vx)$, which an extension of the methodology used by
\cite[][hereafter B14]{Binney2014c} to compute the observables of a
flattened isochrone from an adopted \df\ $f(\vJ)$. 

\subsection{Iterative scheme}

B14 started from  a guess $\Phi_0$ at the self-consistent potential. Using
this guess and equation \eqref{eq:rho_f} he determined the density on a
grid in the meridional $(R,z)$ plane, and then solved Poisson's equation for
the corresponding potential $\Phi_{1/2}(R,z)$. From  this he made an improved
guess 
\[
\Phi_1(R,z)=(1+\alpha)\Phi_{1/2}(R,z)-\alpha\Phi_0(R,z)
\]
at the self-consistent potential, where $\alpha\lta0.5$ is parameter chosen
to maximise the speed of convergence of the sequence of potentials $\Phi_i$
to the self-consistent potential.

The procedure just described involves a grid of points in the $Rz$ plane at
which the density is evaluated by integrating the \df\ over $\vv$. Since the spatial scales of the disc and the
dark halo are very discrepant, it is not cost-effective to use the same grid
for both components: the grid would have to be fine enough to resolve the
$\lta0.1\kpc$ structure of the disc and extensive enough to cover the dark
halo, which has a significant amount of mass $\gta100\kpc$ from the Galactic
centre. Hence for each component $\gamma$ we use a different grid, and on
this grid we store estimates of the density $\rho_\gamma$ of that component
and the contribution $\Phi_\gamma$ that it makes to the total gravitational
potential\footnote{We actually store coefficients of Legendre-polynomial
expansions and store also  radial derivatives.}
\[\label{eq:Phi_sum}
\Phi_{\rm tot}(R,z)=\sum_{\rm components\ \gamma}\Phi_\gamma(R,z).
\]
Evaluation of the density of a single component involves evaluation of the
actions $\vJ(\vx,\vv)$ and in this evaluation we use $\Phi_{\rm tot}$ and
the ``St\"ackel Fudge'' of \cite{Binney2012a}.
Having evaluated the density $\rho_\gamma$ of a component throughout the
component's grid, we solve
Poisson's equation for the corresponding potential $\Phi_\gamma$. The values of
$\Phi_\gamma$ at an arbitrary point $(R,z)$ is obtained by interpolation on
the values stored on the component's grid. 

We start by determining the self-consistent density and potential implied by the
adopted dark-halo \df\ when the halo is isolated. Then our first trial
potential for the whole Galaxy, $\Phi_{\rm tot}$, is the sum of this
potential and the potentials of the bulge, gas disc and double-exponential
stellar disc. We now compute the density of the dark halo in $\Phi_{\rm
tot}$, and solve for the corresponding potential $\Phi_{\rm DM}$ and use this
to update $\Phi_{\rm tot}$. Then we compute the density $\rho_*$ of the
stellar disc in this estimate of $\Phi_{\rm tot}$ and use the corresponding
potential $\Phi_\star$ to update $\Phi_{\rm tot}$. Then we compute the
density of the dark halo in the updated $\Phi_{\rm tot}$, and so on until
successive estimates of $\Phi_{\rm tot}$ differ negligibly. This occurs after
no more than four computations of the density of each component.

\subsection{Reference model}

P14 modelled our Galaxy by combining a spheroidal mass distribution that has
the classic NFW radial profile \citep{NFW1996} and a similar spheroidal bulge
with a fixed gas disc and a dynamical stellar disc that has a \df\ of the
type introduced by \cite{Binney2012b}. The parameters of these components
were adjusted until the composite model reproduced a variety of observational
constraints, such as tangent velocities $v_{\rm los}(l)$ extracted from gas
kinematics at various longitudes, the proper motion of Sgr A* and maser
sources in the disc, and the kinematics of $\sim200\,000$ stars with spectra
taken in the RAdial Velocity Experiment \cite[RAVE][]{RAVE_DR1,RAVE_DR4}.
Crucially, the stellar disc was also required to be consistent with the
density profile above the Sun determined by \cite{Gilmore1983} and
\cite{Juric2008}.

P14 found that the most important unconstrained parameter in their models was
the axis ratio $q\le1$ of the dark halo. The smaller $q$ is, the greater the
fraction of the force on the Sun that comes from dark matter rather than
stars. Since the potentials of the discs cause a dark halo that is spherical
when isolated to flatten in its inner region, we have focused on the model in
P14 that has a mildly flattened dark halo (minor-major axis ratio $q=0.8$).
Moreover, P14 remark that comparison of their results with those of
\citet{Bienayme2014} favours a flattened model with axis ratio $q\simeq0.8$.
We shall refer to this model as our `reference model'.  It was described by
P14 (cf.\ their figures 9, 10 and 11), but they did not specify its
parameters.  Appendix A gives the functional forms used for the density
distributions of the model's components.  Table~\ref{tab:mass_model_params}
contains the values of the parameters that appear in these forms.

Many previous authors have, like P14, assumed that the dark halo will have
the NFW radial profile. But this is the profile of dark halos in \DM\
only simulations. Here we explore the extent to which the dark halo of the
reference model would be modified if the discs determined by P14 were
adiabatically inserted into it.

\section{The DFs} \label{sec:DFs}
\subsection{The stellar disc}
The functional form of the stellar disc's \df, which is made up of
contributions from the thick disc and each coeval cohort of thin-disc stars,
was given by P14 and is reproduced in Appendix B.
Table~\ref{tab:DF_model_params} gives the values of the parameters that
appear in the disc's \df.

We did make one minor change to the functional form of the \df: we imposed a
lower limit, $1\kpc$, on the circular radius $\Rc(J_\phi)$ at which the
epicycle frequencies are evaluated. Implicitly this introduces an upper limit
on the frequencies $\kappa(J_\phi)$, etc., that appear in the \df.  Crucially,
after the limit has been imposed these functions remain well-defined
functions of the actions, so the legitimacy of the \df\ is not undermined.
But on account of the limit, the frequencies employed in the \df\ differ from
the epicycle frequencies at low $J_\phi$.  Imposing the limit has negligible
impact on the structure of the disc near the Sun; its material impact is at
small radii, where the steep rise in the epicycle frequencies as the Galactic
centre is approached would otherwise depress the disc's density in an
unphysical way. Table~\ref{tab:DF_model_params} gives the values of the
parameters in the disc \df.

\subsection{The stellar halo}

P14 used a \df\ $f(\vJ)$
also for the stellar halo, but we omit this component because it has negligible
mass and served only to improve fits to the stellar kinematics at small
$|v_\phi|$, which we do not re-examine here.  

\subsection{The dark halo}

\cite{PontzenGovernato2013} have fitted \df s to dark halos formed in
\DM\ only cosmological simulations. Unfortunately, the \df s they
fitted are not usable here because they have the form $f(E,L)$: a \df\ that
describes one component of a composite model with a self-consistently
generated potential cannot have energy in its argument list for the following
reason.  When components are added, the central potential becomes deeper, so
the energy of the star that rests at a component's centre decreases. If $E$
appears in the argument list, the phase space density around this orbit,
$\vJ=0$, changes in an uncontrolled way. If $f$ increases with $|E|$ as it
generally does, the component's density and mass increases, so $|E|$
increases, and without explicit mass renormalisation, the component's mass is
likely to run away to infinity during successive re-evaluations of the
potential. When, by contrast, $f$ is taken to a function of just the actions, the
phase-space density around each orbit never varies and the iterated
potentials converge rapidly.

In \DM\ only simulations dark halos have NFW density profiles
\citep{NFW1996}.  \cite{Posti2014} found a simple \df\ $f(\vJ)$ that
self-consistently generates a spherical model that has almost exactly an NFW
profile. This model is essentially isotropic near its centre, and becomes
mildly radially biased beyond its scale radius. For the dark halo we adopt
\df s that are based on the form proposed by \cite{Posti2014} even though
these \df s do not generate an NFW profile in the presence of the stellar
disc. The normalisation of the dark-halo \df\ (and thus the halo's total
mass) is chosen to match the circular speed, $v_\mathrm{circ}(R_0)$, in the
reference model at the solar radius $R_0$.  

\begin{figure}
 \centering
 \includegraphics[width=\hsize]{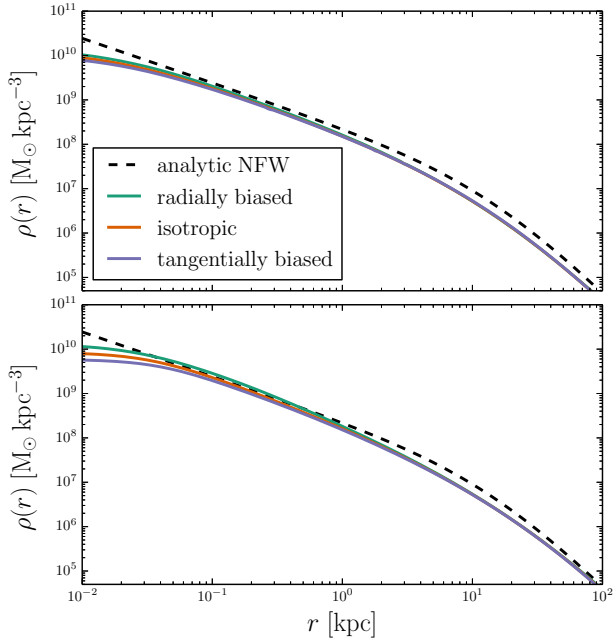}
 \caption{Lower panel: Density profiles computed from the dark halo {\df s}
evaluated in the analytic NFW potential (coloured lines). The associated
analytic NFW density profile is shown as a black dashed line. Upper panel:
density profiles of the isolated halo models after the potential has been
updated to the self-consistently generated one.}
 \label{fig:InitialHaloProfiles}
\end{figure}

We now specify three different
dark-halo \df s that cover the possibilities that the velocity distribution of
\DM\ particles is isotropic or radially or tangentially biased. We
write
\begin{equation}
 f(\vJ) = \frac{N}{J_0^3}\,\frac{[1 + J_0 / ( J_\mathrm{core} + h(\vJ))]^\alpha}{[1 + h(\vJ)/J_0]^\beta} ,
\end{equation}
where $N$ is a normalisation constant, $\alpha$ and $\beta$ are constants
that define the exponents of inner and outer power-law slopes of the density
profile, $J_0$ is a constant that controls the transition between the two.
The constant $J_{\rm core}$ is a small number required to keep the central
density finite.  Finally, 
\begin{equation} \label{eq:hJ}
 h(\vJ)\equiv\frac{1}{A}J_r + \frac{1}{B}\frac{\Omega_\varphi}{\kappa} (|J_\varphi| + J_z)
\end{equation}
is an (approximately) homogeneous function\footnote{\citet{Posti2014} argued
that $h(\vJ)$ should be a homogeneous function of $\vJ$, i.e.\ they required
that $h(a\vJ) = a\,h(\vJ)$.} of $\vJ$ of order unity with
\begin{equation} \label{eq:A}
 A = b_0 + (b - b_0)\tanh^2\left(\frac{|L_z| + J_z}{J_r + |L_z| + J_z}\right),
\end{equation}
\begin{equation} \label{eq:B}
 B = b_0 + (1 - b_0)\tanh^2\left(\frac{|L_z| + J_z}{J_r + |L_z| + J_z}\right)
\end{equation}
and $b_0 = (1+b)/2$. In Appendix C we explain why realistic velocity
distributions are obtained only when the values $A$ and $B$ are functions of
the actions.

The parameter $b$ controls the model's velocity anisotropy. For $b=1$ we
obtain a near-isotropic model, while values below (above) unity yield
tangentially (radially) biased models.  The quantities $\kappa$ and
$\Omega_\varphi$ are epicycle and azimuthal frequencies in the
self-consistent spherical model evaluated at the radius $R_\mathrm{c}$ of a
circular orbit with angular momentum
\[\label{eq:defJtot}
J_{\rm tot}\equiv J_r + |J_\varphi| + J_z.
\]
 We take the argument of $\Rc$ to be $J_{\rm tot}$ to make it an approximate
function of energy, so $R_\mathrm{c}$ does not become small, and the epicycle
frequencies large, for stars on eccentric and/or highly inclined orbits with
small $|J_\varphi|$.

Following \citet{Posti2014}, we set the exponents $\alpha$ and $\beta$ to 5/3
and 2.9, respectively, to obtain a good approximation to the NFW density
profile.  Table~\ref{tab:halo_params} lists all the adopted values of the
parameters that define the halo's \df. They generate a halo that (a) closely
approximates the NFW profile when the halo is isolated, and (b) in the
presence of the discs and bulge has a flattening interior to the Sun similar
to that of the reference model's dark halo.
\begin{table}
 \centering
 \caption{\df\ parameters for the dark halo.}
 \label{tab:halo_params}
 \begin{tabular}{lccc}
 Parameter & Radially biased & Isotropic & Tangentially biased \\
 \hline
 $\alpha$          & 5/3     & 5/3	 & 5/3		\\
 $\beta$           & 2.9     & 2.9	 & 2.9		\\
 $J_0$             & 47.3    & 155	 & 218		\\
 $J_\mathrm{core}$ & 0.00025 & 0.0001	 & 0.0002	\\
 $b$               & 8	     & 1	 & 0.125	\\
  \hline
 \end{tabular}

\end{table}

Once we have obtained a self-consistent model of an isolated halo, we freeze
the dependence of $\Rc$ on its argument, so while we relax $\Phi_{\rm tot}$
onto the potential that is jointly generated by all the Galaxy's components,
the \df\ stays exactly the same function of $\vJ$. It is essential to freeze
the function $f(\vJ)$ during the introduction of the discs and the bulge if
one seeks to learn how the halo is distorted by the gravitational fields of
its companions.
\subsection{The bulge and gas disc}

We could adopt a \df\ $f(\vJ)$ for the final Galactic component, the bulge,
but we do not do so for two reasons.  First, the bulge contains a rotating
bar and at present we cannot represent such an object with a \df\ $f(\vJ)$
\citep[but see][for a non-rotating triaxial example]{SandersBinney2014c}. Second, we are
primarily interested in the structure of the dark halo in the vicinity of the
Sun, which should be well recovered using an axisymmetrised form of the
bulge. Hence we represent the bulge by the fixed density distribution
specified in Appendix A. We also
represent the Galaxy's gas disc by a fixed density distribution.

\section{Results} \label{sec:Results}
\begin{figure}
 \centering
 \includegraphics[width=0.47\textwidth]{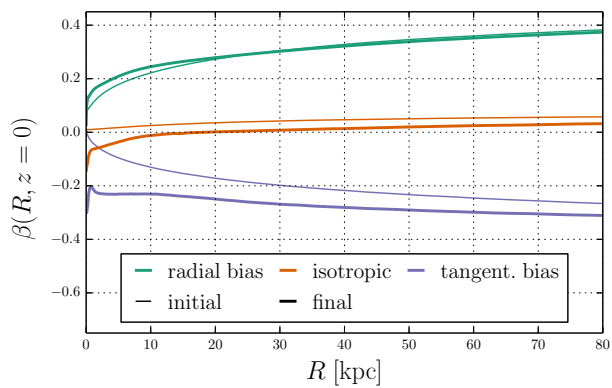}
 \caption{Velocity anisotropy of the \DM\ particles as a function of
radius in the equatorial plane.}
 \label{fig:Halo_Anisotropy}
\end{figure}
\subsection{Isolated halos}
\begin{figure}
 \centering
 \includegraphics[width=0.5\textwidth]{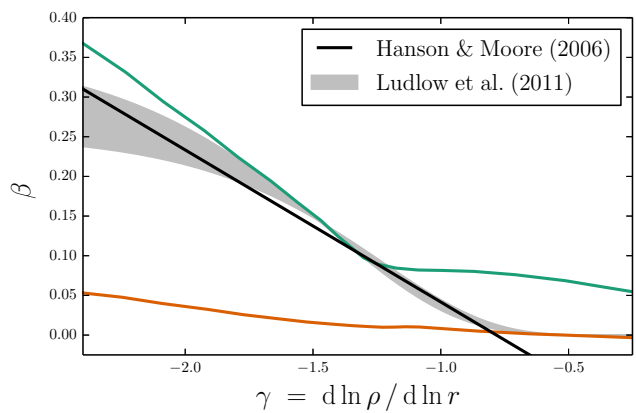}
 \caption{Density slope -- velocity anisotropy relation for our radially
biased (purple) and isotropic models (orange) as well as predictions from
cosmological dark matter-only simulations.}
 \label{fig:GammaBeta}
\end{figure}
We consider three halo \df s, which differ only in the value of the parameter
$b$: we set $b=0.125$ to generate a tangentially biased model, $b=1$ for a
nearly isotropic model, and $b=8$ for a radially biased model. In all plots
we distinguish these models through the colour scheme $b=1$ orange isotropic,
$b=8$ green radially biased, and $b=0.125$ purple tangentially biased.
The lower panel of \figref{fig:InitialHaloProfiles} shows the density profiles
generated by the three \df s in the analytic NFW potential (which has no
core). The \df\ parameters were optimized to provide a good match between
these density profiles and that of the NFW halo fitted by P14. The upper
panel of \figref{fig:InitialHaloProfiles} shows the radial density profiles
of these models after the potential has relaxed to self-consistency. (Fitting
a self-consistent \df-potential pair to the analytic profile directly is
conceptually straightforward, but computationally expensive, and is beyond
the scope of this paper.)

The thin curves in \figref{fig:Halo_Anisotropy} show the anisotropy parameter
\[\label{eq:def_beta}
\beta_{\rm a}\equiv 1 - \frac{\ex{v_{\rm t}^2}}{2\ex{v_r^2}}
\]
for the initial isolated halo as a function of radius in the $xy$ plane.  To
provide some context, \figref{fig:GammaBeta} shows $\beta_{\rm a}$ as a function of
the slope of the logarithmic density profile for two of these models. The
grey band shows the range of such dependency found by \cite{Ludlow2011} from
cosmological simulations, while the black line shows the linear dependence
that \cite{Hansen2006} found described cosmologically simulated halos well.
We see that our isotropic and radially biased models bracket the empirical
results well, except possibly at the smallest values of $|\gamma|$, which
occur at the centre of a halo.
\subsection{Halo distortion}
\begin{figure}
 \centering
 \includegraphics[width=0.47\textwidth]{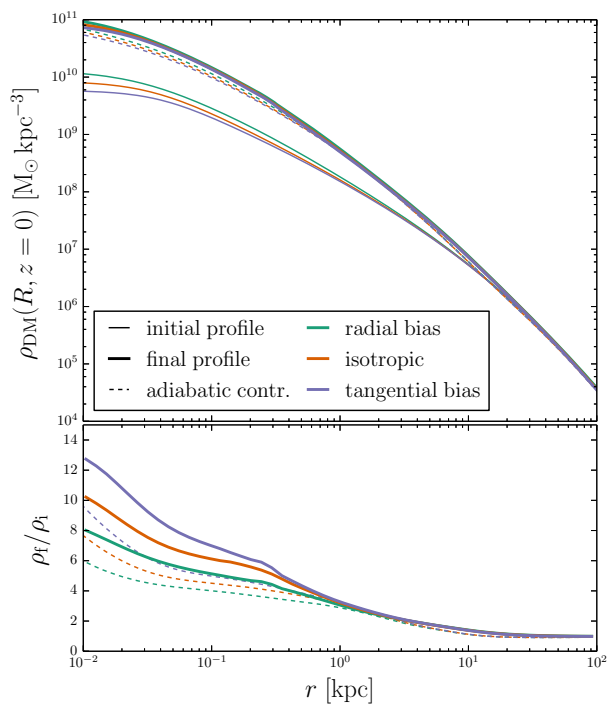}
 \caption{Upper panel: solid thin lines show the spherically symmetric
density profiles of halos in isolation, while thick lines show the density
profiles in the $xy$-plane after the halos have been contracted by
introducing the baryons. The dotted lines show the profiles obtained by
contracting the original models with
the classic adiabatic prescription. Lower panel: final density divided by
initial density when the contraction is done properly (full curves) and using
the classical prescription (dotted curves).}
 \label{fig:HaloProfiles}
\end{figure}
The introduction of the baryonic components causes the initially spherical
dark halo to contract. The end point of this contraction should coincide
perfectly with what we would have found had we slowly grown the baryonic
components in a live N-body simulation of the halo. The almost coincident
heavy curves in the upper panel of \figref{fig:HaloProfiles} show the final,
relaxed \DM\ density profiles in the disc ($xy$) plane.  Even though
the central parts of the initial profiles differ, with the radially-biased
model being most cuspy (light full curves in the upper panel), the final
profiles are very similar. The full curves in the lower panel of
\figref{fig:HaloProfiles} show the ratio of initial to final density at each
radius. We see that in all three cases (radial bias, isotropic, tangential
bias), the central density increases by more than a factor 8.  

This figure shows (i) that at $r\lta3\kpc$ the baryonic component distorts
the dark halo substantially, and (ii) that the variation of the initial
density profiles with the anisotropy parameter $b$ is just what is required
to ensure that the three profiles coincide after the baryons have been added:
as \cite{SellwoodMcGaugh2005} showed, the tangentially biased halo is more
strongly compressed than the radially-biased halo, so before the barons are
added its density needs to be lower than that of the radially-biased halo.

The dotted curves in the upper panel of \figref{fig:HaloProfiles} show the
profiles one obtains when one starts with the three isolated halos and
applies to these objects the classic adiabatic contraction formalism of
\cite{Blumenthal1986}. This formalism is based on the fiction that all dark
matter particles are on circular orbits, i.e. an extreme case of tangential
bias. From  the lower panel of \figref{fig:HaloProfiles} we see that the
analytic prediction is in excellent agreement with our results for $r>1\kpc$,
but under-estimates the contraction for smaller radii.

\begin{figure}
 \centering
 \includegraphics[width=0.5\textwidth]{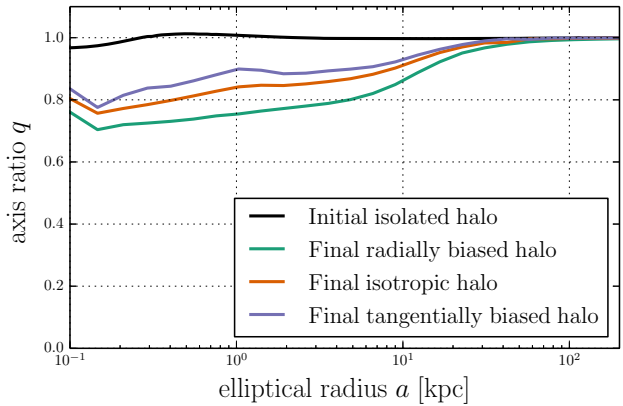}
 \caption{Minor-major axis ratio of the halo component as a function of
elliptical radius $a = \sqrt{R^2 + (z/q)^2}$.}
 \label{fig:Halo_EllipticityProfile}
\end{figure}
We can for the first time study how the flattened baryon distribution breaks
the spherical symmetry of the halo, and pulls it towards the plane --
hitherto analytic prescriptions for adiabatic distortion have been restricted
to the spherical case.  \figref{fig:Halo_EllipticityProfile} shows the axis
ratio of the final dark halo as a function of elliptical radius. We see that
the radially biased halo is most strongly flattened by the baryons -- its
axis ratio does not rise above $0.8$ until near $R_0$. The axis ratio of the
initially nearly isotropic halo rises more gradually from similar values at
$\sim0.2\kpc$ to $\sim0.9$ near the Sun.  We emphasise that these axis ratios
are arising naturally and are \emph{not} related to our choice of a reference
model with a similarly flattened halo. In fact, we found very similar axis
ratios in test runs using the P14 model with a spherical halo and then
inspired by these results chose to use the flattened model as our reference.

\begin{figure*}
 \centering
 \includegraphics[width=0.98\textwidth]{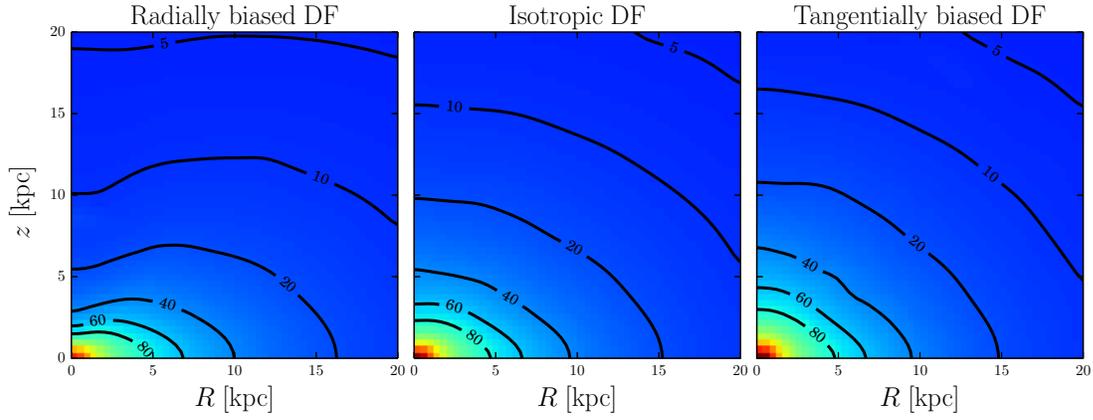}
 \caption{Relative density difference of the dark halo component before and
after the adiabatic contraction, $(\rho_{\rm final}-\rho_{\rm ini})/\rho_{\rm
ini}$, due to the baryonic components. Each panel corresponds to a model with
different velocity anisotropy as indicated above the panels. The contour
lines mark locations where the density has increased similarly by a given
fraction in per cent as indicated on the lines.}
 \label{fig:RelDensMaps}
\end{figure*}
\figref{fig:RelDensMaps} shows the fractional change in the dark halo's
density $(\rho_{\rm final}-\rho_{\rm ini})/\rho_{\rm ini}$ that is induced by
adiabatic insertion of the baryons. We see that for any velocity anisotropy
the effect is strongly concentrated to the inner Galaxy, $r\la5\kpc$. In the
case of radial velocity anisotropy, the dark halo's density is most enhanced
in the region $|z|\la1\kpc$, and the enhancement could be (mis-)interpreted as
the formation of a dark disc around the baryonic disc.
\subsection{Rotation curve}
\begin{figure*}
 \centering
 \includegraphics[width=0.47\textwidth]{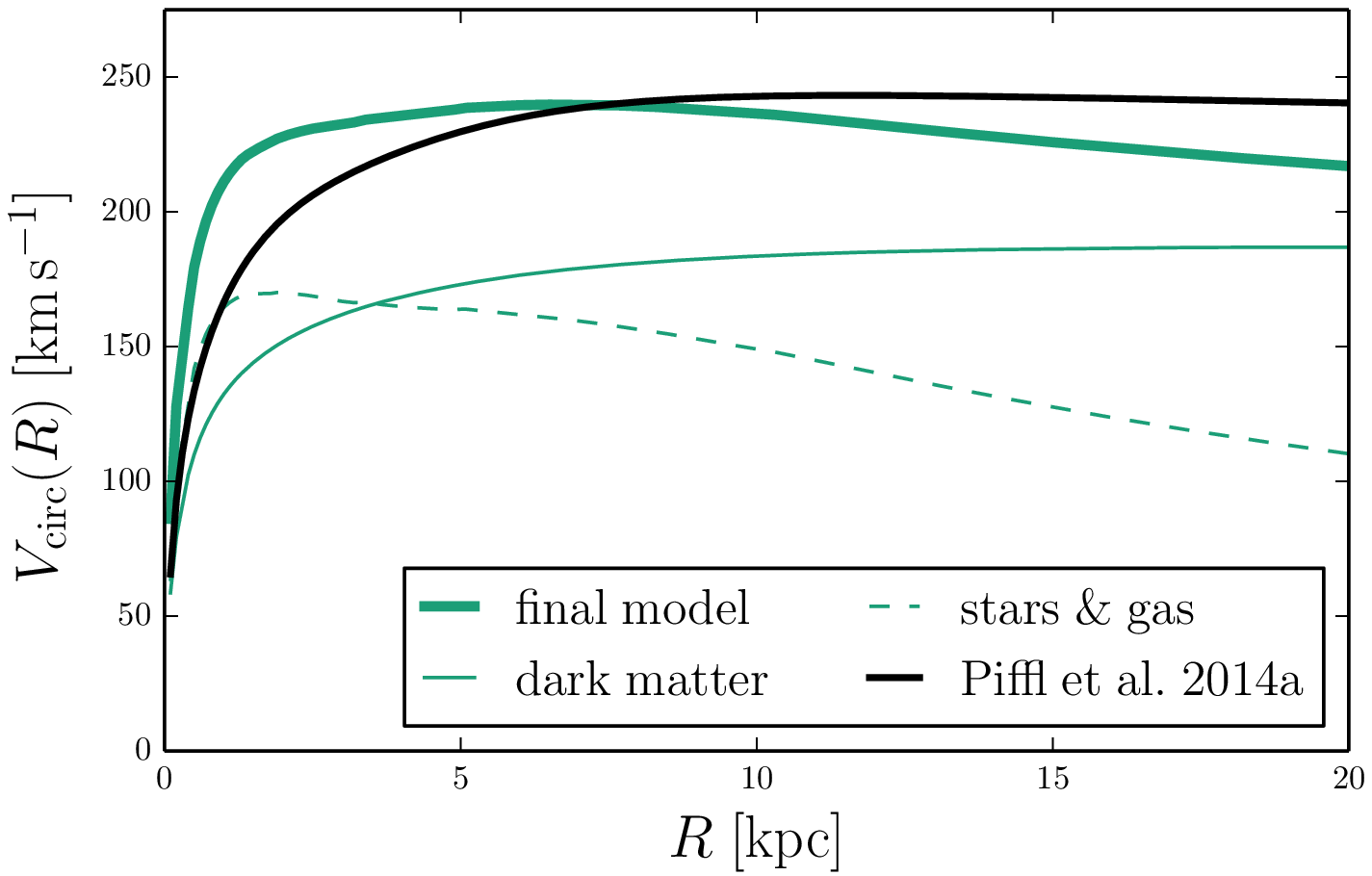}
 \includegraphics[width=0.47\textwidth]{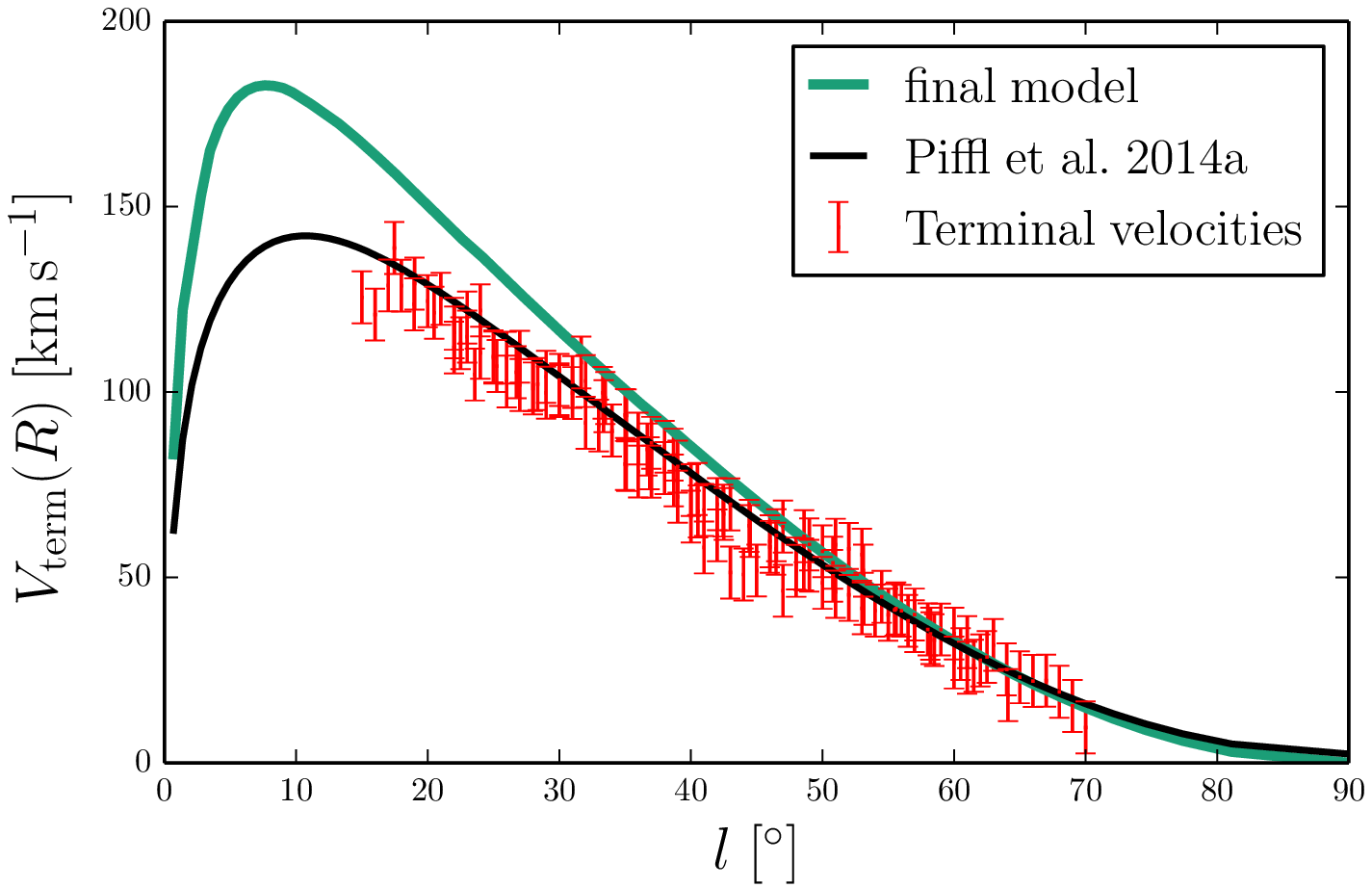}
 \caption{Left: Circular speed curve of the target mass distribution (black
line) and our final self-consistent model with a radially biased velocity
structure in the dark halo. The other two models yield almost
indistinguishable curves. The individual contributions from the dark halo and
the baryonic component are also shown. Right: Terminal velocity curve as a
function of Galactic longitude, $l$, predicted by the same models as in the
left panel and for comparison the HI terminal velocity measurements from
\citet{Malhotra1995} to which the P14 models were fitted.}
 \label{fig:RotationCurve}
\end{figure*}

Our P14 reference model was fitted to data constraining the Galaxy's rotation
curve using an NFW halo profile. As we have seen above, the adiabatic
contraction of the dark halo significantly alters the halo's central density
profile. To counteract these changes a far as possible, we re-scaled the
halo's mass such that we still obtained the right circular speed in the solar
cylinder.  However, because contraction steepens the halo's inner density
profile, the halo's contribution to $v_c^2$ rises inwards faster than in the
P14 model, and when the contracted halo is used in conjunction with the P14
disc and bulge the model no longer matches the constraints on our Galaxy's
inner rotation curve.

Comparison of the black and full green curves in \figref{fig:RotationCurve}
shows the extent of the mismatch. In the left panel, the model circular-speed
curve remains flat down to much lower radii than the curve derived from the
observations. The right panel shows the extent of the conflict with
observations of the terminal velocities of HI gas \citep{Malhotra1995}. This
is a fundamental problem that ties back to the use of an NFW profile in our
reference model. This profile fits baryon-free dark halos and will {\it
not\/} fit dark halos to which baryons have been adiabatically added. If the
addition of baryons is non-adiabatic, the NFW profile {\it
might\/} fit  real dark halos, but we have no reason to suppose that the
addition of baryons was non-adiabatic to the precise extent required for the
NFW profile to remain valid.

\subsection{DM velocity distributions}
We now examine the velocity distribution of the dark matter particles within
the Galactic plane of our models. \figref{fig:Halo_Anisotropy} shows the
radial dependence of the classical anisotropy parameter (equation
\ref{eq:def_beta}) before and after adiabatic contraction.  The most
significant changes are in the innermost regions where the baryonic discs are
dominant.  In the radially biased case (green curves), the anisotropy
increases slightly at $R<15\kpc$ and scarcely changes further out.
In the other two
cases the anisotropy increases at all radii, but no qualitative changes are
observable outside the disc region.

\begin{figure*}
 \centering
 \includegraphics[width=0.98\textwidth]{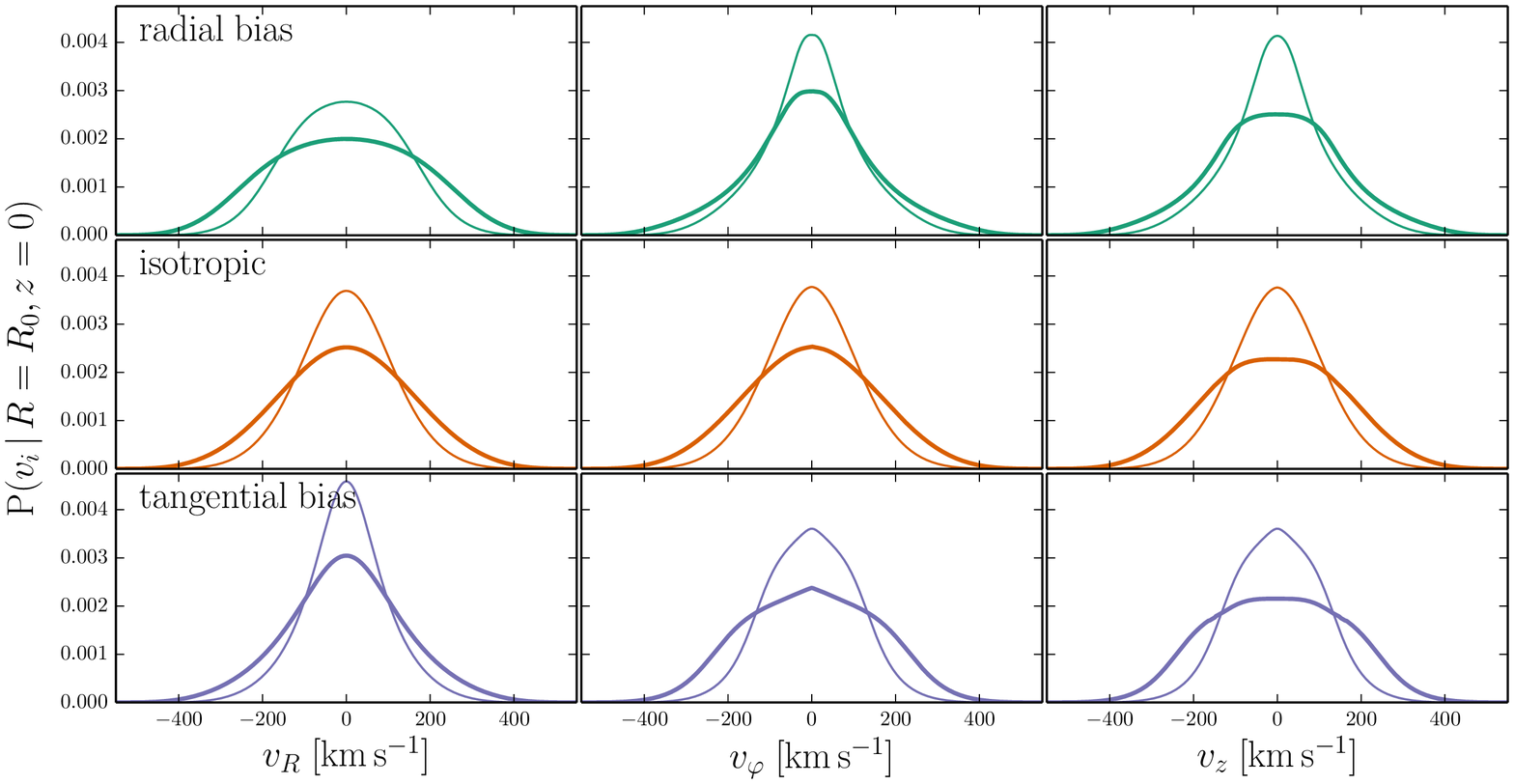}
 \caption{Dark matter velocity distribution in the solar annulus for our
three halo models.}
 \label{fig:Halo_Velocity_dists}
\end{figure*}
\figref{fig:Halo_Velocity_dists} shows the distributions of the three
velocity components (after marginalising over the other two components) of
\DM\ particles at the location $\vx = (R_0,0)$ of the Sun.\footnote{We
adopt $R_0=8.3\kpc$. The distributions at the actual solar
position 14 to $25\pc$ above the Galactic plane are indistinguishable from
these.} The non-Gaussian nature of these velocity distributions is evident.
In all cases adiabatic contraction of the halo broadens the distributions for
the same reason gas that is adiabatically compressed heats up. As the $v_z$-direction broadens, its top becomes remarkably flat.
The dispersions of the distributions are listed in Table~\ref{tab:halo_vel_disps}.
\begin{table}
 \centering
 \caption{Local dark matter velocity dispersions}
 \label{tab:halo_vel_disps}
 \begin{tabular}{lccc}
  bias               & $\sigma_{v_R}$  & $\sigma_{v_\varphi}$ & $\sigma_{v_z}$  \\
                      & $\kms$          & $\kms$               & $\kms$ \\
  \hline
  radial     & 174             & 150                  & 154    \\
  isotropic           & 154             & 154                  & 157    \\
  tangential     & 141             & 156                  & 158    \\
  \hline
 \end{tabular}
\end{table}


\begin{figure}
\includegraphics[width=\hsize]{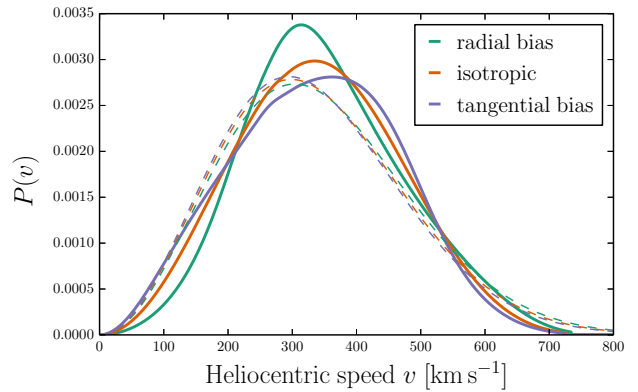}
 \caption{Full curves: the distributions speeds with respect to the Sun of
\DM\ particles at the location of the Sun in our three final models.
Dashed curves show the Maxwellian
distribution with the same dispersions. }\label{fig:Hspeed}
\end{figure}
The sensitivity of particle detectors typically depends on the speed of the
particles to be detected \citep[e.h.][]{Green2012}. The full curves in
\figref{fig:Hspeed} show for each final model the distribution of the speeds
of \DM\ particles with respect to the Sun. The tangentially-biased
model has the most slow-moving and least fast-moving particles, while the
radially biased model has $\sim40\%$ more particles with $v\ga600\kms$. Hence
\DM\ particles will be more readily detected in the radially-biased
case, which is fortunately the case to which N-body simulations point
\citep[e.g.][]{Hansen2006, Ludlow2011, PontzenGovernato2013}.

The
dashed curves in \figref{fig:Hspeed} shows the Maxwellian distributions with
the same velocity dispersions as the real distributions.  These Maxwellian
equivalents are quite similar to one another and differ materially from the
actual distributions. In any stellar system the actual velocity distribution
has to fall below a Maxwellian at large $v$ because it has to vanish above
the escape speed, and this effect is evident in \figref{fig:Hspeed}.

\begin{table}
\caption{Parameters of the best-fitting Tsallis speed distributions shown in
\figref{fig:Tsallis}}\label{tab:Tsallis}
\begin{tabular}{lcc}
bias  &  $q$&  $v_0[\!\kms]$\\
\hline
radial & 0.834 & 341.8\\
isotropic & 0.837 & 332.7\\
tangential & 0.825 & 331.8\\
\hline
\end{tabular}
\end{table}
\begin{figure}
 \includegraphics[width=\hsize]{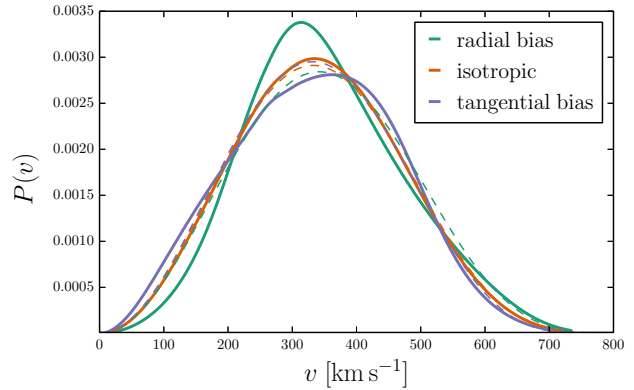}
\caption{The full curves are the same as in \figref{fig:Hspeed}. The dashed
curves show fits to the speed distribution provided by the Tsallis function
(\ref{eq:Tsallis}). Table~\ref{tab:Tsallis} gives the parameters of the
models.}\label{fig:Tsallis}
\end{figure}
A model distribution of speeds that falls to zero at a finite speed is that
proposed by \cite{Tsallis1988}:
\[\label{eq:Tsallis}
n(v)\propto v^2\biggl[1-(1-q){v^2\over v_0^2}\biggr]^{q/(1-q)},
\]
 which tends to a Maxwellian as $q\to1$. \cite{Ling2010} showed that the
Tsallis distribution fitted the speeds of particles in a spherical shell of
radius $r\sim8\kpc$ in their cosmological hydrodynamical simulations of
baryon infall.  The full curves in \figref{fig:Tsallis} are identical to
those in \figref{fig:Hspeed}, while the dashed curves show the best-fitting
Tsallis distributions. Table~\ref{tab:Tsallis} gives the parameters of these
fits.  The fit to the speed distribution of the isotropic (orange) model is
excellent -- in fact, the curve for the Tsallis distribution is frequently
invisible under the true distribution. The fit to the tangentially biased
model is very good, but the Tsallis distribution completely fails to
reproduce the distribution of the radially-biased model.

\subsection{The stellar disc}
\begin{figure}
 \centering
 \includegraphics[width=0.47\textwidth]{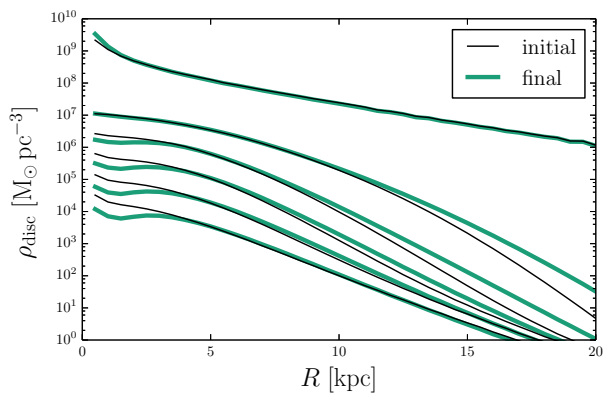}
 \caption{Density profile of the stellar disc in slices of constant $z$ before and after the dark halo's response to the presence of the disc. The line pairs correspond to heights above the Galactic plane from 0 kpc (up-most almost identical lines) to 10 kpc (lowest lines) in steps of 2 kpc.}
 \label{fig:DiscProfile_zSlices}
\end{figure}
\begin{figure}
 \centering
 \includegraphics[width=0.47\textwidth]{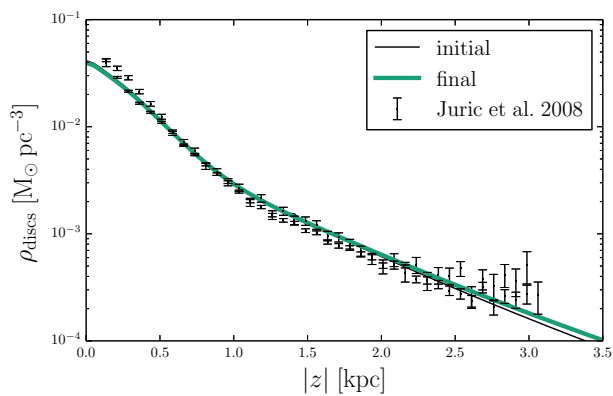}
 \caption{Vertical density profile of the stellar disc in the solar cylinder before and after the response of the halo to the presence of the disc. The star count data from \citet{Juric2008} have been re-normalised (shifted vertically) to allow a comparison to the model.}
 \label{fig:DiscProfile_Sun}
\end{figure}
Figs.~\ref{fig:DiscProfile_zSlices} and \ref{fig:DiscProfile_Sun} show the
stellar density distribution in the case of the radially biased halo by
showing the disc's density at fixed $z$ as a function of $R$ and at $R_0$ as
a function of $z$, respectively. The black curves show the result of
integrating the disc \df\ in the potential of the reference model, while the
green curves are obtained by integrating the same \df\ in the model with the
contracted halo.. The differences between these profiles are significant only
in the outer disc (\figref{fig:DiscProfile_zSlices}) where the final halo is
rounder than that of the reference halo. \figref{fig:DiscProfile_Sun} shows
that in the solar cylinder the two density profiles agree extremely well, so
the final model still provides an excellent match to the star count data of
\citet{Juric2008} to which the reference model was fitted.
\section{Relation to other work}
The growth of baryonic discs in dark halos has been extensively studied with
particle simulations \citep[e.g.][and references therein]{Pillepich2014}. In
the great majority of studies, the manner in which baryons have infiltrated
the dark halo has been controlled by ``sub-grid physics'', that is,
prescriptions governing gas cooling, star formation, and the consequent
heating and ejection of gas. These prescriptions are only loosely based in
physical principles and are freely adjusted to bring certain statistical
properties of the baryonic structures that form into agreement with
observations. In these simulations dark halos do not necessarily evolve
adiabatically -- in fact in most simulations it is likely that scattering of
\DM\ particles by bars and massive gas clouds will have significantly lowered
the peak phase-space density of the dark halo.

\cite{Debattista2008} adiabatically introduced rigid discs into triaxial dark
halos that had previously been formed by colliding spherical dark halos.
After their adiabatic introduction, the discs were adiabatically removed by
evaporation, and the final distributions of dark-matter particles were
compared with the initial distribution. The principal conclusion of this
study was that the initial and final distributions were very similar, which
indicates that the dark halos responded adiabatically. In the absence of a
disc, the halos were mostly strongly prolate, and their outer regions
remained strongly prolate at all times, but the discs distorted the inner
regions of their halos to near axisymmetric oblate forms. Thus
\cite{Debattista2008} underlines the value of being able to compute the
response of a dark halo to the adiabatic insertion of axisymmetric baryonic
structures.

Another study in which for a period the dark halo must have evolved adiabatically
is that of \cite{DeBuhr2012}, for between redshift $z=1.3$ and $z=1$ they
adiabatically grew a massive disc. At $z=1$ the disc became a fully fledged
N-body disc capable of developing a bar and strict adiabaticity ended. The
paper focuses on the axis ratios of the halo and the evolution of the disc.
It provides no quantitative information regarding the evolution of the
phase-space density of \DM\ particles.

\citet{Pillepich2014} compared the final distribution of dark matter in two
simulations of the formation of an object that is slightly less massive than
our Galaxy, in one of which baryons were included while the other contained
only dark matter. The radial DM profiles for the models with and without
baryons in their Figure 1 are very similar to our plots of the DM density
before and after baryon addition in the upper panel of
\figref{fig:HaloProfiles}. They found that the inclusion of baryons increased
\DM\ density at the Sun by $\sim 30$ per cent, in excellent agreement
with our findings. Two processes contribute to enhancement of the \DM\
density in the plane of the simulation with baryons. One is the pinching of
the dark halo by the disc's gravitational field, and the other is the capture
of satellites onto highly inclined orbits through dynamical friction on the
baryonic disc \citep{QuinnGoodman1986} and their subsequent tidal shredding
into a rotating dark disc \citep{Abadi2003b,Read2008, Read2009, Ruchti2014,
Read2014}. Our models include the first mechanism but not the second. The
second process forms a dark disc that rotates at a significant speed in the
same sense as the baryonic disc, while pinching of the dark halo does not set
the halo rotating.  \citet{Pillepich2014} showed that for the rather
quiescent merger history expected for our Galaxy, the effect of pinching
predominates, accounting for $\sim2/3$ of the overall dark disc.

\cite{Ling2010} and \citet{Pillepich2014} used hydrodynamical simulations of
baryon insertion to examine the velocity distribution of \DM\
particles at the Sun's location. The simulation of \citet{Pillepich2014}
involved 13 million DM and 13 million baryonic particles in the
high-resolution region, while the simulations of \cite{Ling2010} employ
slightly fewer particles.  \citet{Pillepich2014} inferred the local velocity
distribution of DM particles from the $\sim80\,000$ DM particles in the
region $|R-R_0|<2\kpc$, $|z|<2\kpc$, while \cite{Ling2010} studied the $2\,662$
particles in the smaller volume $|R-R_0|<1\kpc$, $|z|<1\kpc$. Despite the
enormous computational resources deployed to run these simulations, in
neither study is Poisson noise unimportant, and \citet{Pillepich2014}
considered it necessary to smooth their velocity distribution to a $50\kms$
resolution. Our models require a tiny fraction of the computational resource
and provide velocity distributions that relate precisely to the Sun's
location and yet are completely free of Poisson noise.  Neither
\cite{Ling2010} nor \citet{Pillepich2014} were able to study the dependence
of halo kinematics on the extent of radial bias in the initial dark halo.  

The speed distributions at the Sun reported by \cite{Ling2010} and
\citet{Pillepich2014} are not unlike the one we obtain in the initially
isotropic case. A significant difference between their results and ours is
that they find significant net rotation of the local \DM\ particles
whereas we have none.  For example \cite{Ling2010} find
$\ex{v_\phi}\simeq35\kms$.  Streaming in the dark halo is a consequence of
non-adiabatic interaction between the baryons and the dark halo. In addition
to the shredding of dark halos discussed above, a bar will set a dark halo
spinning by losing angular momentum to it \citep{WeinbergTremaine1984}.

\section{Discussion \& Conclusions}
We have shown how to construct fully self-consistent multi-component galaxy
models with axisymmetric action-based distribution functions $f(\vJ)$.  As a
worked example we have taken a Galaxy model presented by P14 and made its
dark halo a fully dynamical object. In particular, we determined the
gravitational potential that is self-consistently generated by the stellar
disc and the dark halo in the presence of pre-defined contributions from a
gas disc and an axisymmetric bulge. 

To make the dark halo a dynamical object, we must select its velocity
anisotropy. To this end we have introduced a new family of \df s. We have
used the new \df s to
explore three options: radially biased, isotropic and tangentially biased
dark halos. The first two models bracket the predictions of cosmological
\DM\-only simulations.

The most important change in the final model is the adiabatic compression of
the dark halo by the flattened potential of the discs; we can study this
process at a tiny fraction of the computational cost of a cosmological
simulation. 
The contraction has two components. The first component is a spherical
shrinkage.  At radii $r\gtrsim 1\kpc$, i.e., roughly outside one disc scale
height, this is quite well approximated by the simple formalism of
\citet{Blumenthal1986}.  At smaller radii we find stronger contraction. This
is the reverse of what is usually found with N-body simulations, in which
generally less contraction is observed than Blumenthal et al.\ predict
\citep{Abadi2010, Gnedin2010}.

The second component of compression is a pinching towards the plane at radii
$r\lta10\kpc$, which flattens the halo. The extent to which the halo flattens
on introduction of the discs depends of its velocity anisotropy in the sense
that radial bias maximises the flattening. Our initially spherical, radially
biased model deforms to an axis ratio $q< 0.8$ out to $\sim5\kpc$.  The
tangentially biased halo, by contrast, has $q\gtrsim 0.9$ at $r>0.3\kpc$.
This has important consequences, since cosmological simulations predict
radially biased configurations.  Indeed, the left panel of
\figref{fig:RelDensMaps} shows that in the case of a radially biased dark
halo, the difference between the actual, compressed halo and the original
spherical halo looks very much like a ``dark disc''. A ``disc'' of this type
would be non-rotating, in contrast to a dark disc formed by the capture of
satellite halos onto low-inclination orbits followed by tidal shredding
\citep{QuinnGoodman1986, Abadi2003b, Read2008, Ruchti2014, Read2014}.

Our models yield the detailed velocity distribution of \DM\ particles
at the Sun, upon which depends the detectability of dark matter in direct
detection experiments. The distributions of speed with respect to the Sun are
distinctly non-Gaussian, being more and less sharply peaked in the
radially and tangentially biased models, respectively. Given the tendency of
particle detectors to have a threshold speed for detection, the radially
biased model offers significantly better chances of detecting dark matter.

Recently, P14 modelled the Galaxy by combining dynamical models of the
stellar disc with star counts and the kinematics of RAVE stars to obtain
estimates the local \DM\ density. They found that the major model
uncertainty was the flattening of the dark halo.  The best-fitting local
\DM\ density increased with the assumed halo axis ratio $q$ as
$\propto q^{-0.89}$ (cf.\ their Fig.~9). Here we find that even if the dark
halo were originally spherical, the dark halo
must be flattened inside the solar radius $R_0$, with $q = 0.7$ to 0.9. Even
though $q$ varies with radius, this variation is modest within $R_0$, the
relevant region from the perspective of P14.

Using a very different approach to that of P14, \citet{Bienayme2014}
determined the local \DM\ density from a subset of the data used by
P14. The local \DM\ densities of these two studies agree for halo axis
ratio $q$ in the range 0.79 to 0.94. It is interesting that this is just the
range of axis ratios the present study leads us to expect if the dark halo
were spherical before the disc was added.

The \df s of our dark halos were chosen so they generate isolated dark halos
that closely follow the NFW profile. When cohabiting with the P14 disc, the
halos becomes more centrally concentrated, with the consequence that the
final circular-speed curves do not match the data as well as in the reference
model, where the dark halo has an NFW profile in the presence of the discs
and bulge. The models do, however, have appropriate values $v_{\rm c}(R_0)$
for the circular speed at the Sun. 

There are two distinct ways in which we could modify the present model so as
to restore a satisfactory fit to the circular-speed curve while keeping the
dark halo alive. One way is to compensate for the increased central
concentration of the halo by increasing the disc scale lengths from the
values adopted by P14. The other is to modify the \df\ of the dark halo in
such a way that it approximates the NFW profile when its density is evaluated
in the presence of the baryons rather than when isolated. The case for
modification of the \df\ is that the introduction of the baryons was not
completely adiabatic, with the consequence that $f(\vJ)$ decreased near the
origin of action space and increased away from the origin to leave the total
halo mass $(2\pi)^3\int\d^3\vJ\,f(\vJ)$ constant. The modification of the
\df\ could be used to introduce  a degree of rotational streaming such as
will arise from friction against the bar and entrapment of sub-halos on
highly inclined orbits. Cosmological simulations could be used to guide the
choice of halo \df.

While there is a strong case for modifying the halo \df\ from the very simple
form adopted here, it is not clear that the modified \df\ should yield an NFW
profile in the presence of the baryons. Therefore, the next step should be to
see whether a plausible disc \df\ exists which when combined with the present
halo \df\ will produce a satisfactory model of our Galaxy.

We hope soon to report new constraints on the structure of our Galaxy's dark
halo obtained by fitting to observational data model Galaxies in which the
stellar disc, stellar halo and dark halo are all represented by \df s of the
form $f(\vJ)$. The time is also ripe to fit such models to data for external
galaxies, which a new generation of integral-field spectrographs are making
extremely rich. We anticipate that these models will constrain the
\DM\ distributions strongly because the stars will be described by
either distinct \df s for each observationally distinguishable range of
metallicities, or by an ``extended distribution function''
\citep{SandersBinney2015}.
\section*{Acknowledgements}
We thank the member of the Oxford Galactic Dynamics group for helpful
comments and suggestions.

JB was
supported by STFC by grants R22138/GA001 and ST/K00106X/1. 
The research leading to these results has received funding from the European
Research Council under the European Union's Seventh Framework Programme
(FP7/2007-2013)/ERC grant agreement no.\ 321067.
\bibliographystyle{mn2e}
\bibliography{all_references}
\section*{Appendix A: the mass model}
\begin{table}
 \centering
 \caption{Parameters for our reference Galaxy mass model.}
 \label{tab:mass_model_params}
 \begin{tabular}{lcccc}
  \hline
			&			& Gas disc		& Thin disc 		& Thick disc\\[.1cm]
  $\rho_\mathrm{0}$ & [M$_\odot$ kpc$^{-3}$]	& $8.73 \times 10^7$	& $5.32 \times 10^8$ 	& $2.74 \times 10^8$ \\
  $R_\mathrm{d}$ & [kpc]			& $5.16$		& $2.58$		& $2.58$ \\
  $z_\mathrm{d}$ & [kpc]			& 0.04			& 0.20			& 0.67\\
  $R_\mathrm{hole}$ & [kpc]			& 4			& 0			& 0\\[.1cm]
  \hline
 			&			& Bulge			& Dark halo \\[.1cm]
  $\rho_\mathrm{0,b}$ & [M$_\odot$ kpc$^{-3}$]	& $9.49 \times 10^{10}$ & $1.96 \times 10^{7}$ \\
  $r_\mathrm{0,b}$ & [kpc]			& 0.075			& 15.54\\
  $r_\mathrm{cut,b}$ & [kpc]			& 2.1			& $10^5$ \\
  $q_\mathrm{b}$	&			& 0.5 			& 0.8 \\
  $\gamma_\mathrm{b}$	&			& 0			& 1 \\
  $\beta_\mathrm{b}$	&			& 1.8			& 3 \\[.1cm]
  \hline
 \end{tabular}
\end{table}
Our mass models have five components: a gas disc, a thin disc, a thick disc,
a flattened bulge and a dark halo. Since the stellar halo has negligible
mass, we do not include it in the mass model. For the density laws of the
disc components we have
\begin{equation} \label{eq:rho_disc}
 \rho(R,z) = \frac{\Sigma_0}{2z_\mathrm{d}}
    \exp\left[-\left(\frac{R}{R_\mathrm{d}} + \frac{|z|}{z_\mathrm{d}} +
    \frac{R_\mathrm{hole}}{R}\right)\right].
\end{equation}
A non-zero parameter $R_\mathrm{hole}$
creates a central cavity in the disc. This is used to model
the gas disc while for the other two discs $R_\mathrm{hole}$ is set to zero. The values of
the parameters are given in Table~\ref{tab:mass_model_params}.

The density distributions of  the dark halo and the bulge components are
\begin{equation} \label{eq:rho_spheres}
 \rho(R,z) = \frac{\rho_0}{m^\gamma(1+m)^{\beta-\gamma}}
 \exp[-(mr_0/r_\mathrm{cut})^2],
\end{equation}
where
\begin{equation}
 m(R,z) = \sqrt{(R/r_0)^2 + (z/qr_0)^2}.
\end{equation}
The parameters of the reference model are given in
Table~\ref{tab:mass_model_params}.

\section*{Appendix B: the disc DF}
The disc \df\ is built up out of ``quasi-isothermal'' components. The \df\ of
such a component is
\[\label{eq:qi}
 f(J_r,J_z,L_z)=f_{\sigma_r}(J_r,L_z)f_{\sigma_z}(J_z,L_z),
\] 
where $f_{\sigma_r}$ and $f_{\sigma_z}$ are defined to be
\[\label{planeDF}
 f_{\sigma_r}(J_r,L_z)\equiv \frac{\Omega\Sigma}{\pi\sigma_r^2\kappa}
 [1+\tanh(L_z/L_0)]\e^{-\kappa J_r/\sigma_r^2}
\]
and
\[\label{basicvert}
 f_{\sigma_z}(J_z,L_z)\equiv\frac{\nu}{2\pi\sigma_z^2}\,
 \e^{-\nu J_z/\sigma_z^2}.
\]
Here $\Omega(L_z)$, $\kappa(L_z)$ and $\nu(L_z)$ are, respectively, the
circular, radial and vertical epicycle frequencies of the circular orbit with
angular momentum $L_z$, while
\[\label{eq:defsSigma}
 \Sigma(L_z)=\Sigma_0\e^{-\Rc/\Rd},
\]
where $\Rc(L_z)$ is the radius of the circular orbit, determines the surface
density of the disc: to a moderate approximation this surface density can be
obtained by using for $L_z$ in equation (\ref{eq:defsSigma}) the angular
momentum $L_z(R)$ of the circular orbit with radius $R$. The functions
$\sigma_r(L_z)$ and $\sigma_z(L_z)$ control the radial and vertical velocity
dispersions in the disc and are approximately equal to them at $\Rc$. Given
that the scale heights of galactic discs do not vary strongly with radius
\citep{vanderKruit1981}, these quantities must increase inwards. We adopt the
dependence on $L_z$
\begin{eqnarray}
 \sigma_r(L_z)&=&\sigma_{r0}\,\e^{(R_0-\Rc)/R_{\sigma,r}}\cr
 \sigma_z(L_z)&=&\sigma_{z0}\,\e^{(R_0-\Rc)/R_{\sigma,z}},
\end{eqnarray}
so the radial scale-lengths on which the velocity dispersions
decline are $R_{\sigma,i}$. Our expectation is that $R_{\sigma,i} \sim 2\Rd$.

In equation (\ref{planeDF}) the factor containing tanh serves to eliminate
retrograde stars; the value of $L_0$ controls the radius within which
significant numbers of retrograde stars are found, and should be no larger
than the circular angular momentum at the half-light radius of the bulge.
Provided this condition is satisfied, the results for the extended solar
neighbourhood presented here are essentially independent of $L_0$.

We take the \df\ of the thick disc to be a single pseudo-isothermal. The thin
disc is treated as a superposition of the cohorts of stars that have age
$\tau$ for ages that vary from zero up to the age $\tau_{\rm max} \simeq
10\Gyr$ of the thin disc. We take the \df\ of each such cohort to be a
pseudo-isothermal with velocity-dispersion parameters $\sigma_{r}$ and
$\sigma_{z}$ that depend on age as well as on $L_z$.  Specifically, from
\citet{Aumer2009} we adopt
\begin{eqnarray}\label{sigofLtau}
 \sigma_r(L_z,\tau)&=&\sigma_{r0}\left(\frac{\tau+\tau_1}{\tau_{\rm m}+\tau_1}\right)^\beta\e^{(R_0-\Rc)/R_{\sigma,r}}\nonumber\\
 \sigma_z(L_z,\tau)&=&\sigma_{z0}\left(\frac{\tau+\tau_1}{\tau_{\rm m}+\tau_1}\right)^\beta\e^{(R_0-\Rc)/R_{\sigma,z}}.
\end{eqnarray}
Here $\sigma_{z0}$ is the approximate vertical velocity dispersion of local
stars at age $\tau_{\rm m}\simeq10\Gyr$, $\tau_1$ sets velocity dispersion at
birth, and $\beta\simeq0.33$ is an index that determines how the velocity
dispersions grow with age. We further assume that the star-formation rate in
the thin disc has decreased exponentially with time, with characteristic time
scale $t_0$, so the thin-disc \df\ is
\[\label{thinDF}
 f_{\rm thn}(J_r,J_z,L_z)=\frac{\int_0^{\tau_{\rm m}}\rd\tau\,\e^{\tau/t_0}
 f_{\sigma_r}(J_r,L_z)f_{\sigma_z}(J_z,L_z)}{t_0(\e^{\tau_{\rm m}/t_0}-1)},
\]
where $\sigma_r$ and $\sigma_z$ depend on $L_z$ and $\tau$ through equation
(\ref{sigofLtau}). We set the normalising constant $\Sigma_0$ that appears in
equation (\ref{eq:defsSigma}) to be the same for both discs and use for the
complete \df
\[
 f_{\rm disc}(J_r,J_z,L_z)=f_{\rm thn}(J_r,J_z,L_z) +
 F_{\rm thk}f_{\rm thk}(J_r,J_z,L_z),
\]
where $F_{\rm thk}$ is a parameter that controls the fraction $(1+F_{\rm
thk}^{-1})^{-1}$ of stars that belong to the thick disc.\footnote{Note, that $F_{\rm thk}$
is the ratio of the total masses of the thick and the thin discs, while the
parameter $\fthick$ used for the mass model is the ratio of the local mass
densities of the two discs. Hence the two parameters are intimately related
but not the same.} The values of the parameters are given in
Table~\ref{tab:DF_model_params}.

\begin{table}
 \centering
 \caption{Parameters for stellar disc \df.}
 \label{tab:DF_model_params}
 \begin{tabular}{lcccc}
  \hline
			&		& Thin disc 	& Thick disc \\[.1cm]
  $\sigma_\mathrm{r}$	& $\kms$ 	& 33.9		& 50.5 \\
  $\sigma_\mathrm{z}$	& $\kms$ 	& 25.0		& 48.9 \\
  $R_\mathrm{d}$	& $\kpc$	& 2.58		& 2.58 \\
  $R_{\sigma,r}$	& $\kpc$	& 9		& 12.9 \\
  $R_{\sigma,z}$	& $\kpc$	& 9		& 4.1 \\
  $\tau_1$		& Gyr		& 0.01		& - \\
  $\tau_{\rm m}$	& Gyr		& 10		& - \\
  $\tau_0$		& Gyr		& 8		& - \\
  $\beta$		& -		&0.33		& - \\
  $F_\mathrm{thick}$	& -		& -		& 0.455 \\[.1cm]
  \hline
 \end{tabular}
\end{table}

\section*{Appendix C: cusps and dimples in $v_\phi$ distributions}
\begin{figure}
 \centering
 \includegraphics[width=0.47\textwidth]{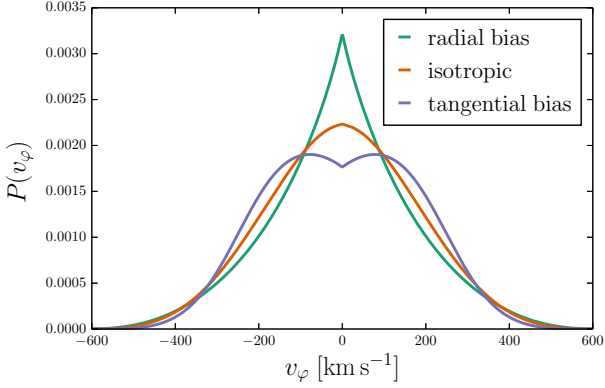}
 \caption{Local $v_\varphi$ distributions using a simpler form of the halo \df\ with a constant anisotropy parameter.}
 \label{fig:oldVphiDistributions}
\end{figure}
To implement velocity anisotropy in our halo \df\ we did not opt for the
simplest possible solution. Here we explain the reasons for this decision. In
the \df\ the two factors $A$ and $B$ (Eqs.~\ref{eq:A}, \ref{eq:B}) control
the anisotropy of the system by enhancing or suppressing the population of
orbits with certain properties. These factors are constant except for very
radial orbits. This variation makes $h(\vJ)$ not a strictly homogeneous
function any more, a property that was demanded by \citet{Posti2014} when
they proposed this class of \df s. 

If we remove the $\vJ$ dependency by substituting the $\tanh$ function with
unity we obtain $A=b$ and $B=1$ which is sufficient to create an anisotropic
model. However, when we examine the tangential velocity distribution of such
a model (\figref{fig:oldVphiDistributions}) we find that in the radially
biased model, the distribution in $v_\varphi$ has a cusp at $v_\phi=0$, while
in the tangentially biased model it has a dimple at $v_\phi=0$; only the
isotropic model has a $v_\phi$ distribution of the expected domed shape. The
cusp and dimple become more pronounced after adiabatic contraction. 

The cusp and the dimple arise because the energy, upon which the ergodic \df\
of the isotropic model depends, is a quadratic function of $v_\phi$, while
$J_\phi=Rv_\phi$ is a linear function of $v_\phi$. So when the \df\ of the
isotropic model is written as a function of the actions and then Taylor
expanded in $v_\phi$ at fixed $\vx$, the linear dependence of $f$ upon $v_\phi$ that arises
from the dependence of $f$ on $J_\phi$ has to be cancelled by linear
dependencies of $J_r$ and $J_z$ on $v_\phi$. If these linear dependencies do
not cancel, $\partial f/\partial v_\phi$ is non-zero at $v_\phi=0$, and
$P_{v_\phi}$ acquires a cusp or a dimple there. Below we show that
this delicate cancellation is disturbed by shifting $b$ from unity.

The derivative of the $v_\varphi$ distribution $P_{v_\varphi}$ plotted in
\figref{fig:Halo_Velocity_dists} is
\begin{equation} \label{eq:dPdv}
 \frac{\d P_{v_\varphi}}{\d v_\varphi}(v_\varphi\,|\,\vx) 
= \iint \d v_R \d v_z \frac{\partial f}{\partial v_\varphi}\bigg|_{\vx}\,,
\end{equation}
and
\begin{equation} \label{eq:dfdv}
 \frac{\partial f}{\partial v_\varphi}\bigg|_{\vx} 
= \left(\frac{\partial f}{\partial J_r}\frac{\partial J_r}{\partial v_\varphi} +
  \frac{\partial f}{\partial J_\varphi}\frac{\partial J_\varphi}{\partial v_\varphi} +
  \frac{\partial f}{\partial J_z}\frac{\partial J_z}{\partial
  v_\varphi}\right)_{\vx}.
\end{equation}
Now
\begin{equation}
 \frac{\partial f}{\partial J_r} = \frac{\partial f}{\partial h(\vJ)} \frac{1}{b}
\end{equation}
and
\begin{equation}
 \frac{\partial f}{\partial J_\varphi} = \frac{\partial f}{\partial J_z} = \frac{\partial f}{\partial h(\vJ)} \frac{\Omega_\varphi}{\kappa}
\end{equation}
We require  derivatives with respect to $v_\phi$ at a fixed location
$\vx=(R,0)$ in the plane, so
\begin{equation}
 \frac{\partial J_\varphi}{\partial v_\varphi} = R\,,
\end{equation}
where we have used $J_\varphi = R v_\varphi$.

We don't have analytic expressions for $\partial J_r/\partial v_\varphi$ and
$\partial J_z/\partial v_\varphi$, but the following argument shows that
$\partial J_r/\partial v_\varphi<0$ for $0\le v_\varphi\ll v_\mathrm{circ}(R)$.
Consider a star with low $v_\varphi$, and therefore a small guiding radius,
that is just able to reach $(R,0)$, i.e.\ has its apocentre there. It is on
an eccentric orbit with substantial $J_r$.  Increasing $v_\varphi$ moves the
guiding centre outwards and thus diminishes the amplitude of the star's
radial oscillations and lowers its value of $J_r$.

To see that $\partial J_z/\partial v_\varphi|_{\vx}<0$ for $0\le v_\varphi\ll
v_\mathrm{circ}(R)$, consider the case of a spherical potential. Then
$J_z=L-|J_\phi|$ with $L=R\sqrt{v_\phi^2+v_z^2}$ in the Galactic plane, so
\[
\frac{\partial J_z}{\partial v_\phi}=\frac{\partial L}{\partial v_\phi}-R
=\frac{R^2v_\phi}{L}-R=R\left(\frac{J_\phi}{L}-1\right),
\]
 which is clearly negative for small $v_\phi$.
In a strongly flattened potential, the vertical motion largely decouples from
the planar motion, so $\partial J_z/\partial v_\phi$ becomes numerically
small, but it remains negative.

Inserting all these insights into equation \eqref{eq:dfdv} we arrive at
\begin{equation} \label{eq:dfdv2}
 \frac{\partial f}{\partial v_\varphi} 
= \frac{\partial f}{\partial h(\vJ)}
\left(\frac{1}{b}\frac{\partial J_r}{\partial v_\varphi} 
+ \frac{\Omega_\varphi}{\kappa}R 
+ \frac{\Omega_\varphi}{\kappa}\frac{\partial J_z}{\partial v_\varphi}\right)
\end{equation}
for the integrand in equation \eqref{eq:dPdv}. Since $\partial f / \partial
h(\vJ)<0$, the sign of $\d P_{v_\varphi} / \d v_\varphi$ is opposite to the
sign of the bracket in equation \eqref{eq:dfdv2}. In the isotropic case ($b=1$) we
must have $\d P_{v_\varphi} / \d v_\varphi \simeq 0$ at $v_\varphi = 0$, so
\begin{equation} \label{eq:balance}
 \frac{\partial J_r}{\partial v_\varphi} + \frac{\Omega_\varphi}{\kappa} \frac{\partial J_z}{\partial v_\varphi} \simeq -\frac{\Omega_\varphi}{\kappa} R\,.
\end{equation}
Hence for $b\ne1$ this balance is disturbed with the consequence that
$P_{v_\phi}$ acquires a non-zero gradient at $v_\phi=0$, the gradient being
negative when $b>1$ and positive when $b<1$ as is evident in
\figref{fig:oldVphiDistributions}.

While cusps and dimples are not strictly unphysical, we think it unlikely
that real \DM\ distributions would exhibit them. A requirement to
avoid these features amounts to a restriction on permissible forms of the
function $h(\vJ)$ as, e.g.\ the one we choose for this work.
\label{lastpage}
\end{document}